\documentclass[%
 reprint,
 amsmath,amssymb,
 aps,
twocolumn
]{revtex4}

\newcommand{\lsim}   {\mathrel{\mathop{\kern 0pt \rlap
  {\raise.2ex\hbox{$<$}}}
  \lower.9ex\hbox{\kern-.190em $\sim$}}}
\newcommand{\gsim}   {\mathrel{\mathop{\kern 0pt \rlap
  {\raise.2ex\hbox{$>$}}}
  \lower.9ex\hbox{\kern-.190em $\sim$}}}

\usepackage{graphicx}
\usepackage{dcolumn}
\usepackage{bm}
\usepackage[utf8]{inputenc}
\usepackage{slashed}
\usepackage{physics}
\usepackage[colorlinks=true,linktoc=all]{hyperref}
\usepackage{xcolor}
\usepackage{changebar}

\DeclareMathOperator{\diag}{diag}

\begin{document}

\preprint{}

\title{The imprint of ultralight vector fields on gravitational-wave propagation}

\author{Alfredo D. Miravet}
 \email{alfrdelg@ucm.es}
\affiliation{Departamento de F\'{\i}sica Te\'orica and Instituto de F\'isica de Part\'iculas y del Cosmos (IPARCOS), Universidad Complutense de Madrid, 28040 Madrid, Spain}
\author{Antonio L. Maroto}
\email{maroto@ucm.es}
\affiliation{Departamento de F\'{\i}sica Te\'orica and Instituto de F\'isica de Part\'iculas y del Cosmos (IPARCOS), Universidad Complutense de Madrid, 28040 Madrid, Spain}


\date{\today}

\begin{abstract}
We study the effects of ultralight vector field (ULVF) dark matter on gravitational-wave propagation. We find that the coherent oscillations of the vector field induce an anisotropic suppression of the gravitational-wave amplitude as compared to the  $\Lambda$CDM prediction. The effect is enhanced for smaller vector field masses and peaks for modes around $k=H_0/\sqrt{a(H=m)}$. The suppression is negligible for astrophysically generated gravitational waves but could be sizeable for primordial gravity waves. 
We discuss the possibility of detecting such an effect on the tensor  power spectrum with future CMB B-mode polarization detectors.  We find that for the sensitivity of the upcoming LiteBIRD mission, the correction to the tensor power spectrum at decoupling time could be distinguishable from that of $\Lambda$CDM for ULVF masses  $m\lsim 10^{-26}$ eV and sufficiently large abundances.
\end{abstract}

\maketitle

\section{Introduction}\label{sec:Intro}

In recent years there has been a growing interest in the so-called ultralight dark matter models. These models, also known as  fuzzy dark matter, are based on the existence of bosonic fields with very small masses $m\ll 1$ eV, so that their cosmological  number density is so high that the interparticle separation would be smaller than their Compton wavelength. In this case, the appropriate description of the dark matter component would be in terms of classical fields (Bose-Einstein condensates) rather than particles. These models could alleviate some of the small-scale problems of the cold dark matter scenario and lead to distinctive predictions for the   density profiles of galaxies. Thus, recent simulations of dark matter in the form of a Bose-Einstein condensate have shown the formation of a soliton core profile on scales below the de Broglie wavelength surrounded by a halo of excited states \cite{Schive:2014dra}.

Although ultralight scalars, i.e. spin-zero fields with very small masses ($m\sim 10^{-22}$ eV) such as axion-like particles, have been amongst the most studied proposals \cite{Hui:2016ltb,Arias:2012az, Cembranos:2018ulm}, fuzzy dark matter can be extended to higher-spin fields. In that sense, ultralight vector fields (ULVF) \cite{Cembranos:2016ugq,Agrawal:2018vin,Co:2018lka,Bastero-Gil:2018uel,Dror:2018pdh,Nomura:2019cvc}, spin-1 bosons with very small masses ($m\ll 1$ eV) and very weak interactions, have been growing in popularity throughout the last years, also making a good candidate for dark matter. The effect of ULVF on the scalar sector of cosmological metric perturbations has already been analysed in \cite{Cembranos:2016ugq}, and it produces the desired effect of suppressing structure formation on scales smaller than its comoving de Broglie wavelength $\lambda_{dB}=(\mathcal{H}ma)^{-1/2}$. The typical structure size in our Universe is associated with a field of mass $m\sim10^{-22}$ eV, meaning that lighter fields have an astrophysical-sized de Broglie wavelength, and thus could not conform the totality of the dark matter in the Universe.

Aside from its dynamics, its production has also been studied in different contexts, such as inflation \cite{Graham:2015rva, Nakai:2020cfw} or by misalignment mechanisms \cite{Nelson:2011sf}.  However, it has been shown that even though the standard misalignment mechanism works for a wide mass range, it requires large, highly tuned non-minimal couplings to the scalar curvature \cite{Arias:2012az}. In addition, these couplings lead to perturbative unitarity violation at low energies in longitudinal photon-graviton scattering, and also to a negative kinetic term of the longitudinal mode for a certain range of momenta, thus threatening the vacuum stability. These issues have been discussed in previous works but none of them have succeeded in ensuring the viability of the mechanism \cite{AlonsoAlvarez:2019cgw}. As a result, alternative mechanisms which include couplings to axion fields have been proposed in \cite{Agrawal:2018vin,Co:2018lka,Bastero-Gil:2018uel,Dror:2018pdh}. On the other hand, the possibility of having coherently oscillating spin-2 fields as dark matter (DM) candidates has also been explored in \cite{Cembranos:2013cba,Aoki:2017cnz}.

Regarding detection of ULVF DM, the possible indirect detection through the generation of induced atomic transitions was considered in \cite{Alvarez-Luna:2018jsb}, the use of optomechanical accelerometers as resonant detectors was explored in \cite{Manley:2020mjq}. The generation of gravitational waves (GWs) associated to density perturbations was studied in \cite{Cembranos:2016ugq}. It was found that the generated signal would be very small to be detected by gravitational-wave interferometers and could only have an effect on the CMB. More recently, possible effects on pulsar timing signals have been considered in \cite{LopezNacir:2018epg,Nomura:2019cvc}.

On the other hand, the detection and confirmation of gravitational waves a few years ago \cite{Abbott:2016blz} has opened a new window in observational cosmology. Although all observations to date have their source in stellar-mass astrophysical  objects, namely black holes \cite{Abbott:2016blz} and neutron stars \cite{TheLIGOScientific:2017qsa}, cosmological primordial GWs,  as predicted by most inflationary models, are expected to exist and to be observable. The future detection of primordial GW could take place not directly but through the measurement of the low-multipole ($\ell < 200$) region of the CMB polarization B-mode angular power spectrum (see \cite{Kamionkowski:2015yta} for a review), which is expected to be measured with enough precision in the coming years \cite{Hui:2018cvg,Suzuki:2015zzg,Hazumi:2019lys}. Having travelled all the way through from the end of inflation, primordial GW propagation is very sensitive of any modification of matter-energy content of the universe in any cosmological era. In particular and unlike  perfect fluids or scalar fields, the presence of a coherent vector field that could play the role of dark matter, induces a non-zero contribution to the tensor anisotropic stress, which enters the GW  equation via the metric perturbation of the stress-energy tensor, and modifies their propagation. 


The aim of this work is precisely to analyse the effects of ultralight vector fields on GW propagation. With that purpose, we consider a model based on a
homogeneous massive abelian vector field. We analyse the parameter space in which the field behaves as ULVF dark matter and its dynamics and equation of state in past epochs. Then, we study its effects on GW propagation and calculate the range of modes and parameters for which the impact is larger with respect to standard $\Lambda$CDM. Finally, we discuss the regimes in which this field could result in a non-zero change to the predicted CMB B-mode observation.

The work is organised as follows. Firstly, in Section \ref{sec:GWprop} we review the foundations of GW propagation in vacuum. In Section \ref{sec:vfdynamics} we present our model of ULVF, get the governing equations of its dynamics and discuss its validity as dark matter. Sections \ref{sec:perturbations} and \ref{sec:gwabundance} are dedicated to obtaining the GW propagation equations and the GW abundance expression in the presence of the vector field and to a qualitative analysis of the solutions. In Section \ref{sec:numresults} we introduce the numerical model to solve the equations, present the different results and make some discussion about them. Finally, in Section \ref{sec:conclusions} we draw the main conclusions of the work.

\section{GW propagation in vacuum}\label{sec:GWprop}
In this section, we derive the equation for gravitational-wave propagation in vacuum. The metric we consider is  flat  Robertson-Walker (RW) in conformal time $\eta$, which up to first order in metric perturbations can be written as
\begin{equation}\label{eq:RW}
	g_{\mu\nu}=a^2(\eta)(\eta_{\mu\nu}+\delta\eta_{\mu\nu}),
\end{equation}
where $\eta_{\mu\nu}=\diag(+,-,-,-)$. The equation to solve is the perturbed Einstein equation in vacuum
\begin{equation}
	\delta G^\mu{}_\nu=0,
\end{equation}
where $\delta G$ is first order in metric perturbations. In order to do so, we perform a helicity decomposition on the first-order tensor $\delta\eta_{\mu\nu}$ in order to separate scalar, vector and tensor contributions, and we are left with the following components:
\begin{subequations}
	\begin{equation}
		\delta\eta_{00}=2\psi,
	\end{equation}
	\begin{equation}
		\delta\eta_{0i}=\beta_i-\partial_i\gamma,
	\end{equation}
	\begin{equation}
		\delta\eta_{ij}=2\phi\delta_{ij}-\left(\partial_i\partial_j-\frac{1}{3}\delta_{ij}\nabla^2\right)\lambda-\frac{1}{2}(\partial_i\epsilon_j+\partial_j\epsilon_i)-h_{ij}^{TT}.
	\end{equation}
\end{subequations}

In all these equations, latin indices run over spatial components $i,j=1,2,3$. Out of all these quantities, we are interested solely in its transverse-traceless (TT) contribution $h_{ij}^{TT}$, which is the only tensor in the helicity decomposition, and so it suffices to describe gravitational waves. From now on, the TT indicators will be dropped. $h_{ij}$ is gauge invariant, and also symmetric, transverse and traceless, as all scalar and vectorial behaviours have been removed from it, so it satisfies
\begin{equation}
	h_{ii}=\delta^{ij}h_{ij}=0,\qquad \partial_i h_{ij}=0.
\end{equation}

Thus, all scalars and vectors can be dropped for our purpose, and from now on we will use:
\begin{equation}
    \delta\eta_{00}=0,\qquad \delta\eta_{0i}=0,\qquad \delta\eta_{ij}=-h_{ij}.
\end{equation}

Notice that in perturbations, spatial indices are raised and lowered with $\delta_{ij}$. Note how, since the inverse metric is given by
\begin{equation}
	g^{\mu\nu}=a^{-2}(\eta)(\eta^{\mu\nu}-\delta\eta^{\mu\nu}+\mathcal{O}(\delta\eta^2)),
\end{equation}
we have $\delta\eta^{ij}=-h^{ij}=-\delta^{ik}\delta^{jl}h_{kl}$.

With all this, the perturbed Einstein tensor can be computed, whose only non-zero components are
\begin{equation}\label{eq:deltaGij}
	\delta G^i{}_j=-\frac{1}{2a^2}\left(h_{ij}''+2\mathcal{H}h_{ij}'-\nabla^2h_{ij}\right),
\end{equation}
where $'\equiv \dv*{\eta}$ and $\mathcal{H}=a'/a$.

\begin{figure}
	\includegraphics{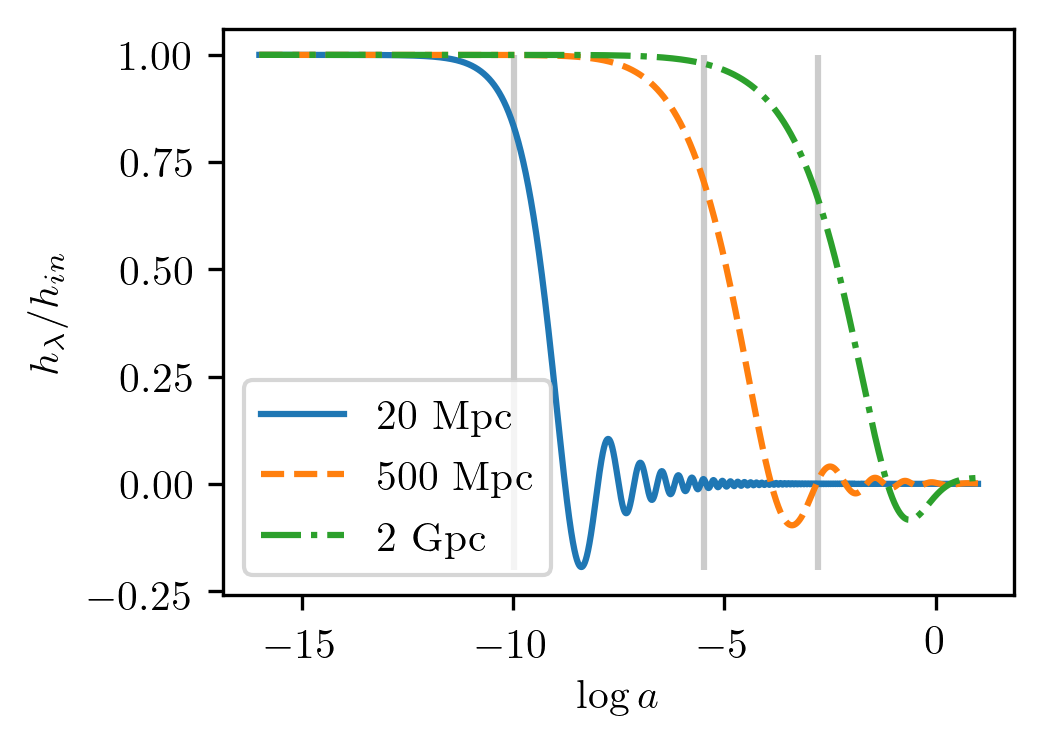}
	\caption{Time evolution of GWs in $\Lambda$CDM, normalised by their primordial value, with three different reduced wavelengths $\lambdabar=1/k$ as indicated in the legend. The vertical lines indicate when each GW enters the Hubble horizon, i.e. $k=\mathcal{H}$, and coincides with the moment the damped oscillation starts.}
	\label{fig:hvac}
\end{figure}

If we choose positive $i=3$ to be the direction of propagation of the wave, defined by the momentum vector $\vb{k}$, $h_{ij}$ can be written in terms of the two possible polarizations
\begin{equation}\label{eq:hijmatrix}
	h_{ij}=\begin{pmatrix}
		h_+ & h_\times & 0\\
		h_\times & -h_+ & 0\\
		0 & 0 & 0
	\end{pmatrix},
\end{equation}
and after going through a Fourier transformation, we are left with two identical equations
\begin{equation}
	h_{\lambda}''+2\mathcal{H}h_{\lambda}'+k^2h_{\lambda}=0,\qquad \lambda=+,\times,
\end{equation}
which is the equation that governs the GW propagation in vacuum. The evolution of an arbitrary GW mode is qualitatively simple, and depends solely on the wavelength of the GW: If the mode is outside the Hubble horizon $k\ll\mathcal{H}$, it remains constant, whereas if it is inside the Hubble horizon $k\gg\mathcal{H}$, it oscillates with the amplitude damped as $1/a$. An example of the evolution for different wavelengths is depicted in Fig. \ref{fig:hvac}.

\section{Dynamics of a homogeneous vector field in cosmology}\label{sec:vfdynamics}

Let us consider a massive abelian vector field \cite{Cembranos:2013cba} in a Robertson-Walker (RW) background. The corresponding action reads
\begin{equation}\label{eq:SA2}
	S=\int\dd{x}\sqrt{g}\left(-\frac{1}{4}F_{\mu\nu}F^{\mu\nu}+\frac{1}{2}m^2A_\mu A^\mu\right),
\end{equation}
where $g=|\det(g_{\mu\nu})|$, $F_{\mu\nu}=\partial_\mu A_\nu-\partial_\nu A_\mu$, and the minus sign in the potential is due to stability requirements, since the vector field is going to be purely spacelike, as we are about to see.

The equations of motion for the vector field are obtained by varying the action \eqref{eq:SA2} with respect to $A_\mu$, yielding
\begin{equation}\label{eq:A2eom}
	F^{\mu\nu}{}_{;\nu}-m^2 A^\mu=0,
\end{equation}
where the semicolon represents covariant derivative.

We consider a homogeneous vector field, thus dependent solely on conformal time $\eta$,  whose spatial components we choose, for simplicitly, to  point in a fixed direction (linear polarization). After conveniently orienting the spatial axes, it can be written as
\begin{equation}\label{eq:Amuansatz}
	A_\mu(\eta)=(A_0(\eta),0,0,A_z(\eta)),
\end{equation}
and since we are working at the background level,  the metric \eqref{eq:RW} with $\delta \eta_{\mu\nu}=0$ is enough to compute the equations of motion. Fixing $\mu=0$ in \eqref{eq:A2eom} gives
\begin{equation}
	m^2 A_0=0,
\end{equation}
so the temporal component vanishes, whereas the spatial part of the equation yields
\begin{equation}\label{eq:A2zeom}
	A_z''+m^2a^2A_z=0.
\end{equation}

We can study now the behaviour of $A_z$ at different  stages of the cosmological evolution. It can be seen that for $ma\gg\mathcal{H}$ the field oscillates rapidly around the potential minimum. Thus, introducing a WKB ansatz
\begin{equation}
	A_z(\eta)=F_k(\eta)\cos\int^\eta\omega(\eta')\dd{\eta'}
\end{equation}
and substituting back in \eqref{eq:A2zeom} up to  next-to-leading adiabatic order, we get
\begin{equation}\label{eq:wkbsol}
	A_z(\eta)=A_{z,0}a^{-1/2}(\eta)\cos\int^\eta ma(\eta')\dd{\eta'},
\end{equation}
where $A_{z,0}$ is a normalization constant.

Now, in order to study the effect of this vector field in the RW background, we need to compute the stress-energy tensor $T^\mu{}_\nu$ for this theory. This can be obtained by varying the action with respect to the metric tensor, so that
\begin{equation}\label{eq:Tdef}
	\delta S=-\frac{1}{2}\int \dd{x}\sqrt{g}\,T^{\mu\nu}\delta g_{\mu\nu}.	
\end{equation}

After varying the action \eqref{eq:SA2}, we obtain
\begin{equation}\label{eq:TA2}
	T^\mu{}_\nu=\left(\frac{1}{4}F_{\rho\sigma}F^{\rho\sigma}-\frac{1}{2}m^2A_\rho A^\rho\right)\delta^\mu{}_\nu-F^{\mu\rho}F_{\nu\rho}+m^2A^\mu A_\nu,
\end{equation}
Notice that since the background vector field breaks isotropy, it cannot be the source of a RW background. However as shown in \cite{Cembranos:2012kk,Cembranos:2013cba}, for sufficiently fast oscillations, the average energy-momentum tensor is always isotropic. Thus, we can write down the Friedmann equation with the stress-energy tensor given by this theory
\begin{equation}
		\mathcal{H}^2=\left(\frac{a'}{a}\right)^2=\frac{8\pi G}{3}\rho_A a^2,
\end{equation}
with
\begin{equation}\label{eq:rhoA}
	\rho_A=T^0{}_0=\frac{A_z'^2}{2a^4}+\frac{m^2}{2a^2}A_z^2.
\end{equation}
If, instead, we want to consider this ULVF within the framework of $\Lambda$CDM, we must simply add this energy density to those from all the other species when writing the Friedmann equation.

\begin{figure}
	\includegraphics[width=\linewidth]{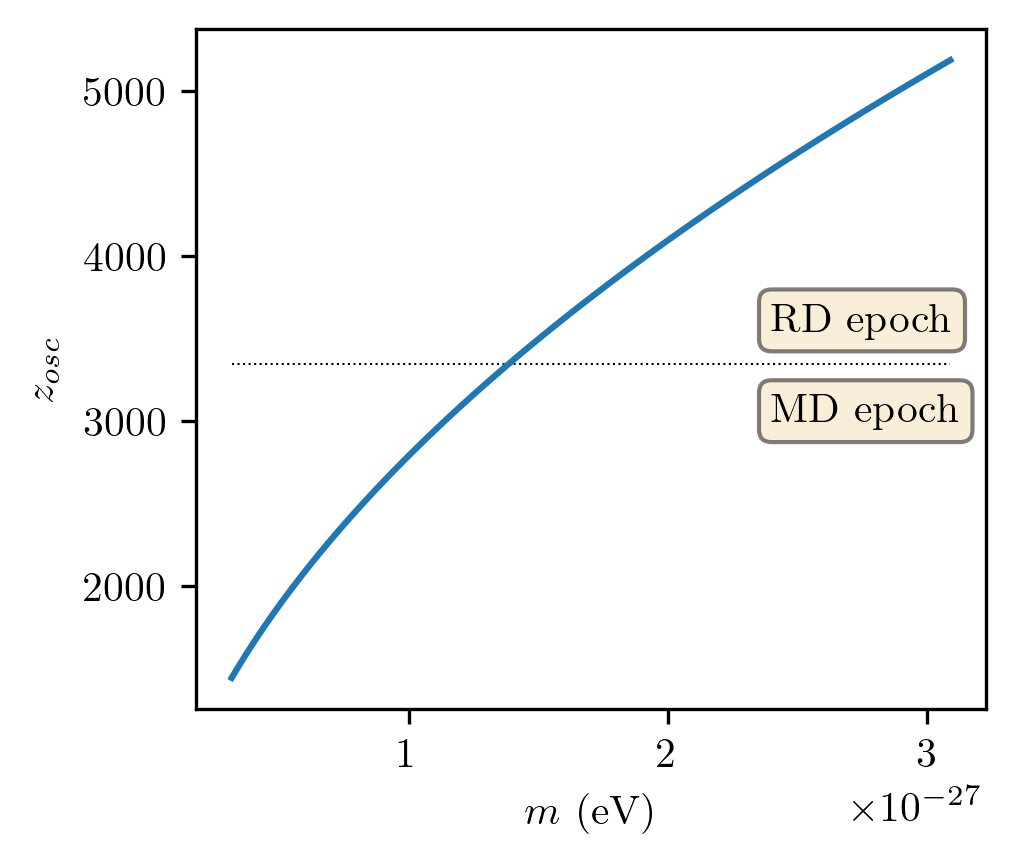}
	\caption{Starting point of vector field oscillation expressed in redshift as a function of its mass, computed with a $\Lambda$CDM background. The horizontal line corresponds to the matter-radiation equality redshift $z_{eq}\approx 3350$, which separates radiation-dominated (RD) and matter-dominated (MD) epochs. Consistency with $\Lambda$CDM requires this moment to occur in RD epoch, yielding a lower bound for the ULVF mass.}
	\label{fig:mdrd}
\end{figure}

With the aid of the WKB solution \eqref{eq:wkbsol} and Eq. \eqref{eq:rhoA}, we can see that $\rho_A\propto a^{-3}$, which means that, in the regime in which the WKB condition $ma\gg\mathcal{H}$ is valid, the vector field behaves as matter, so that it could be part of the unexplained dark matter component. Consequently, we must ensure this behaviour throughout the whole matter-dominated epoch, since the matter-radiation equality must remain unaffected. Fig. \ref{fig:mdrd} displays the redshift at which the field starts to oscillate as a function of its mass, from which we get the constraint $m\gtrsim  10^{-27}$ eV.

\begin{figure}
	\includegraphics[width=\linewidth]{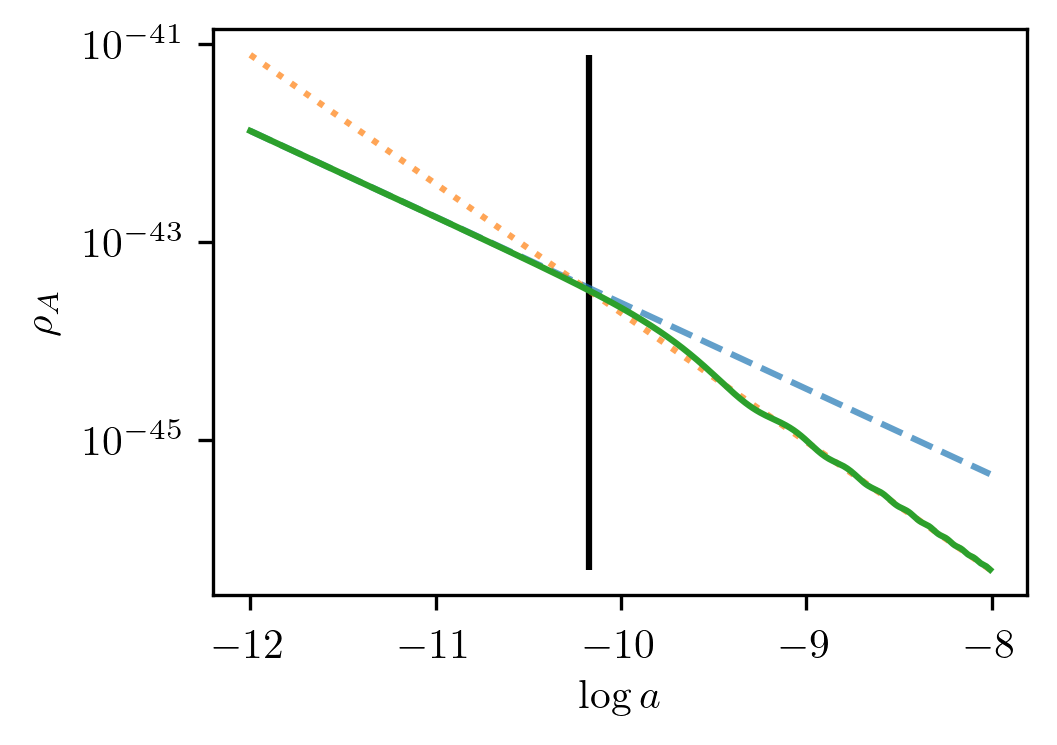}
	\caption{ULVF energy density (in arbitrary units) as a function of the scale factor, for $m=10^{-26}$ eV in a $\Lambda$CDM background. Dashed and dotted lines represent $a^{-2}$ and $a^{-3}$ (matter-like) scalings respectively. The vertical line corresponds to $ma=\mathcal{H}$, after which the energy density oscillates around $\rho_A\propto a^{-3}$ as a consequence of the field oscillation.}
	\label{fig:rhoA}
\end{figure}

At earlier times when the WKB condition is not satisfied, if one imposes that the time derivative of the field is zero at early epochs, it remains constant at its initial value $A_{z,0}$, so that the energy density scales as $\rho_A\propto a^{-2}$. This holds until $ma\simeq\mathcal{H}$ is reached, when it begins to oscillate as described before and behaves as matter, with the energy density oscillating around the $a^{-3}$ scaling, as can be seen in Fig. \ref{fig:rhoA}.

\section{Introducing metric tensor perturbations}\label{sec:perturbations}
The effect of the presence of a vector field in the GW propagation can be studied through the perturbed Einstein equation, now with a non-zero stress-energy tensor. In addition, since we are interested only in TT modes, we need to project the Einstein equation onto these modes, which reads
\begin{equation}
	\Lambda_{ij,lm}(\delta G^l{}_m-8\pi G\delta T^l{}_m)=0,
\end{equation}
where
\begin{equation}
	\Lambda_{ij,lm}=P_{il}P_{jm}-\frac{1}{2}P_{ij} P_{lm}
\end{equation}
is the TT projector and
\begin{equation}
	P_{ij}=\delta_{ij}-\hat{k}_i\hat{k}_j.
\end{equation}

The transverse traceless projection of the perturbed Einstein tensor yields exactly \eqref{eq:deltaGij}, as $\Lambda_{ij,lm}h_{lm}=h_{ij}$, because this tensor is already transverse and traceless. On the other hand, after perturbing the metrics in the stress-energy tensor \eqref{eq:TA2}, omitting the $\delta^\mu{}_\nu$ term as it projects to zero, and doing the projection, one gets
\begin{equation}
	\delta T^i{}_j= -\frac{1}{a^4}h_{ik}A'_jA'_k+\frac{m^2}{a^2}h_{ik}A_jA_k,
\end{equation}
so that
\begin{widetext}
\begin{eqnarray}
	\Lambda_{ij,lm}\delta T^l{}_m=-\frac{1}{a^4}\left[h_{ik}	A'_jA'_k-\hat k_j\hat k_m h_{ik}A'_mA'_k-\frac{1}{2}(\delta_{ij}-\hat k_i \hat k_j)h_{lk} A'_lA'_k\right]\nonumber\\
	+\frac{m^2}{a^2}\left[h_{ik}	A_jA_k-\hat k_j\hat k_m h_{ik}A_mA_k-\frac{1}{2}(\delta_{ij}-\hat k_i \hat k_j)h_{lk} A_lA_k\right].
\end{eqnarray}
\end{widetext}

Notice that because of the presence of the background vector field, it would be possible to have tensor contributions at the linear level coming from scalar and vector modes \cite{Cembranos:2016ugq} that would source gravitational waves. However, in this work we will only concentrate on the effect on propagation so that we will ignore the new source terms.   

We calculate the components of these tensors in the orthonormal basis \cite{Cembranos:2016ugq}  defined by the vectors $\{\vu{u}_1,\vu{ u}_2, \vu{ u}_3\}=\{\vu{ u}_{pk}, \cos\theta \vu{ u}_p-\sin \theta \vu{ u}_a,\sin\theta \vu{ u}_p+\cos\theta\vu{u}_a\}$, where $\vu{u}_a$ is the unit vector that points in the direction of the vector field, which remains constant during its evolution, so that $\vb{A}=A\vu{u}_a$, $\vu{u}_3=\vu{k}$ is the direction of propagation of the gravitational wave, $\cos\theta=\vu{k}\vdot\vu{u}_a$ and
\begin{equation}
	\vu{u}_{pk}=\frac{\vu{k}\cp\vu{u}_a}{\sin\theta},\qquad \vu{u}_p=\frac{\vu{k}-\cos\theta\vu{u}_a}{\sin\theta}.
\end{equation}

Since $\vu{u}_3$ is the GW direction of propagation, we can use the expression \eqref{eq:hijmatrix} for $h_{ij}$. Thus, in this basis, we easily get the following expression from the Einstein equation projection:
\begin{equation}\label{eq:h2eq}
	h_{\lambda}''+2\mathcal{H}h_{\lambda}'+\left[k^2-8\pi G\sin^2\theta\left( \frac{A'{}^2}{2a^2}-\frac{m^2}{2}A^2 \right) \right]h_{\lambda}=0,
\end{equation}
with $\lambda=+,\times$.

If the vector field is not oscillating (i.e. it remains constant) when the GW mode enters the Hubble horizon, the only effect we effectively get is a shift in the momentum $k$ and thus a slight displacement of the instant when the mode enters the horizon. On the other hand, more interesting effects, such as changes in the GW amplitude, are expected to happen to modes that enter the horizon when the vector field is already oscillating, in particular to all modes that become sub-Hubble during the matter-dominated epoch.

Let us write the vector field abundance today, when the WKB approximation must be valid, as
\begin{equation}\label{eq:wA}
	\Omega_A=\frac{\rho_{A,0}}{\rho_c}=\frac{4\pi G}{3}\frac{m^2A_{z,0}^2 }{H_0^2},
\end{equation}
where $H_0$ is the Hubble parameter in cosmological time $H=\mathcal{H}/a$ today and $\rho_c=3H_0^2/8\pi G$ is the critical energy density today.  Note that, as this field behaves as matter today, we must have $\Omega_A<\Omega_M$ necessarily. If we introduce this as well as the WKB expression for the vector field in \eqref{eq:h2eq} and we perform a change of variable $h_\lambda=v_\lambda/a$, we get the following equation:
\begin{widetext}
\begin{equation}
	v_\lambda''+\left[k^2-\frac{a''}{a}+\frac{3\Omega_A H_0^2\sin^2\theta}{a}
\cos\left(2\int^\eta ma(\eta')\dd{\eta'}\right)\right]v_\lambda=0.
\end{equation}
\end{widetext}
So that modifications in the GW propagation could appear given the following requirements:
\begin{itemize}
	\item   The vector field must be oscillating at some point of the cosmological evolution: $ma\gg\mathcal{H}$, which is ensured as long as we keep the matter-radiation equality unaffected, as discussed before.
	\item   The vector field term must be greater or at least of the order of the $k^2$ and the damping terms, i.e.  $3\Omega_A H_0^2\sin^2\theta/a\gtrsim k^2,$ $a''/a$.
	\item   If the vector field term oscillates very quickly, it could be averaged out, resulting in no effect. Thus, the frequency $ma$ cannot be much larger than the GW oscillation frequency, and since the former is monotonically increasing, the effect is going to be more noticeable if the vector field has oscillated for a short time when the GW mode enters the horizon.
\end{itemize}

Taking all of this into account, the most affected modes are going to be those around $k^2=H_0^2/a_*$, with $a_*$ defined as the scale factor at which the vector field starts behaving as matter, which can be approximately determined by $ma_*=\mathcal{H}(a_*)$. The complicated structure of the differential equation does not allow much further analysis, and all calculations and results must be obtained numerically.

The comparison between having this vector field or not in the GW evolution can be done through a \emph{quotient function} that will depend on all parameters in our model. Let $|h|$ be the oscillation amplitude of the GW if it is inside the horizon or the (constant) value itself if outside the horizon, then we can define
\begin{equation}\label{eq:Qfun}
	Q(k,m,\theta,\Omega_A)=\frac{|h_{\lambda,\Lambda\text{CDM+A}}(k,m,\theta,\Omega_A)|}{|h_{\lambda,\Lambda\text{CDM}}(k)|}.
\end{equation}
as the ratio between the GW amplitudes in the $\Lambda$CDM model with an additional ULVF dark matter component and the standard $\Lambda$CDM, assuming the same initial conditions. 
For sufficiently long times, the quotient function is independent of the time at which we evaluate both solutions, since they are both constant or equally damped as $1/a$. Notice that it is also independent of the GW polarization. 

\section{Power spectrum and GW abundance}\label{sec:gwabundance}

Another interesting quantity to compute, as it is widely used to express the sensitivity of GW detectors, is the gravitational-wave abundance today per energy interval \cite{Maggiore:1900zz}. Typically it is integrated over solid angle, but the anisotropy of our model in the vector-GW angle $\theta$ makes this integration non-trivial, so the quantity to compute is
\begin{equation}\label{eq:omegagw1}
	\dv{\Omega_{GW}}{\cos\theta}=\frac{1}{\rho_c}\dv{{}^2\rho_{GW}}{\log k \dd{\cos\theta}},
\end{equation}
where the gravitational-wave density today is given by
\begin{equation}
	\rho_{GW}=\frac{\langle h_{ij}'h_{ij}'\rangle}{32\pi G}.
\end{equation}

This can be related to the tensor power spectrum at redshift $z$,  $\mathcal{P}_{T}(z,k,\theta)$, and can ultimately be written in terms of the primordial tensor power spectrum $\mathcal{P}_{T,in}$ through the transfer function for $\Lambda$CDM $T(k,z)$ 
\begin{equation}\label{eq:Tfun}
	h_{\lambda,\Lambda\text{CDM}}(k,z)=T(k,z) h_{\lambda,\Lambda\text{CDM}}(k,\eta_{\text{in}})
\end{equation}
and the quotient function. The transfer function relates the amplitude of gravity waves of a certain wavelength to its  primordial value in $\Lambda$CDM, and the quotient function accounts for the difference between $\Lambda$CDM and our model. Power spectra relate via squared transfer functions, so that
\begin{equation}
	\mathcal{P}_{T}(z,k,\theta)=\frac{1}{2}|T(k,z)|^2|Q(k,m,\theta,\Omega_A)|^2\mathcal{P}_{T,in}(k).
\end{equation}
where the $1/2$ factor is needed to keep the correct normalization when the angular integration is performed. 
The primordial spectrum is usually parametrized as
\begin{equation}
	\mathcal{P}_{T,in}(k)=A_T(k_*)\left(\frac{k}{k_*}\right)^{n_T},
\end{equation}
where $k_*$ is the pivot scale, $n_T$ is the spectral index and $A_T$ is the spectrum amplitude at the pivot scale. Finally, we can write the later in terms of the comoving curvature power spectrum amplitude $A_\mathcal{R}$ and the tensor-to-scalar ratio $r$:
\begin{equation}
	r(k_*)=\frac{A_T(k_*)}{A_\mathcal{R}(k_*)}.
\end{equation}

With all of this combined, we can finally write down 
\begin{eqnarray}\label{eq:omegagw2}
	\mathcal{P}_{T}(z,k,\theta)=\frac{1}{2}r(k_*)A_\mathcal{R}(k_*)\left(\frac{k}{k_*}\right)^{n_T}\nonumber\\
	|T(k,z)|^2|Q(k,m,\theta,\Omega_A)|^2.
\end{eqnarray}
and for the GW abundance
\begin{eqnarray}
	\dv{\Omega_{GW}}{\cos\theta}=\frac{k^2}{12H_0^2}\mathcal{P}_{T}(z,k,\theta)
\end{eqnarray}
Thus, the only calculations within our model that we must perform is the determination of the quotient function.

\section{Numerical model and results}\label{sec:numresults}
Conformal time $\eta$ does not appear explicitly in any of the equations, so we can get rid of it to work in terms of the scale factor $a(\eta)$, which is more convenient. However, as the scale factor spans several orders of magnitude, we will work with the $x=\log a(\eta)$ variable in order to improve calculation performance. Eqs. \eqref{eq:A2zeom} and \eqref{eq:h2eq}, to solve in terms of $x$, are listed as follows:
\begin{subequations}
	\begin{equation}\label{eq:A2x}
		\partial_x^2 A+\frac{\partial_x\mathcal{H}}{\mathcal{H}}\partial_x A +\frac{m^2 e^{2x}}{\mathcal{H}^2}A=0,
	\end{equation}
	\begin{eqnarray}\label{eq:hl2x}
		\partial_x^2 h_\lambda&+&\left(2+\frac{\partial_x\mathcal{H}}{\mathcal{H}}\right)\partial_x h_\lambda+\frac{1}{\mathcal{H}^2}\bigg[k^2\\
		&-&4\pi G\sin^2\theta\left(e^{-2x}\mathcal{H}^2(\partial_x A)^2-m^2A^2\right)\bigg]h_\lambda=0.
		\nonumber
	\end{eqnarray}	
\end{subequations}

The inconvenience of eliminating conformal time is that now the Hubble parameter appears in the equations. Its expression  \emph{including} the vector field energy density can be written as
\begin{eqnarray}\label{eq:H2x}
	\mathcal{H}^2=H_0^2 e^{2x}\bigg[(\Omega_M-\Omega_A)e^{-3x}+\Omega_R e^{-4x}+\Omega_\Lambda\nonumber\\
	+\frac{4\pi G}{3H_0^2e^{2x}}\left(\frac{\mathcal{H}^2(\partial_x A)^2}{e^{2x}}+m^2A^2\right)\bigg],
\end{eqnarray}
where the last term in the square brackets must equal $\Omega_A$ when evaluated today ($x=0$). The matter abundance today consisting of the ultralight vector field is subtracted from the total matter $\Omega_M$, yielding the term $(\Omega_M-\Omega_A)a^{-3}$ which includes the CDM and baryon contributions. Notice that depending on the $\Omega_A$ value, the share of ULVF dark matter vs. CDM varies.

The way of solving this system would be to solve \eqref{eq:H2x} for $\mathcal{H}$. Then,  plugging the result into \eqref{eq:A2x}, we get the solution for the vector field. Finally we solve the GW equation \eqref{eq:hl2x}. This procedure, however, has several inconveniences. The main one is using the expression for $\mathcal{H}$ that contains $A$ and its derivative explicitly, as it enlarges considerably the equation for the vector field. In addition, having a denominator that can be close to zero  results in numerical inaccuracies. On the other hand, large masses for the vector field translate into thousands of oscillations before arriving to the present moment $x=0$, which can take long computing times even if the $\Lambda$CDM Hubble parameter is used. Lastly, we would like $\Omega_A$ to be an input parameter of our model, but since it is a quantity calculated at $x=0$ rather than an initial value, this requirement would turn our equations into a boundary value problem, in which the primordial value of the vector field should be fine-tuned to yield the desired $\Omega_A$, with a shooting method for example, making the process fairly slow.

Instead, we can argue that the effect of the vector field on the background does not have a large impact on its own evolution. This is sensible as the vector energy density scales initially as $a^{-2}$ and then behaves as matter in a radiation-dominated epoch, making it subdominant throughout this period. After a few oscillations of the field, the WKB approximation is valid and its presence can be accounted by treating it as standard matter. Thus, the equation for the vector field \eqref{eq:A2x} is integrated using the standard $\Lambda$CDM Hubble parameter
\begin{equation}\label{eq:Hlcdm}
	\mathcal{H}_{\Lambda\text{CDM}}^2=H_0^2 e^{2x}\left(\Omega_M e^{-3x}+\Omega_R e^{-4x}+\Omega_\Lambda\right),
\end{equation}
with initial conditions $A(x_{\text{in}})=A_{\text{in}}$, $A'(x_{\text{in}})=0$, until the scale factor grows by a factor $e^2$ after the field started oscillating. At this point, it  oscillates sufficiently fast as to use the WKB solution, so we match both solutions at the first minimum after this point, which satisfies $A(x_{\text{min}})/A_{\text{in}}=A_{\text{min}}$, and with that we can write
\begin{equation}
	A(x)=A_{\text{in}}A_{\text{min}} e^{-(x-x_{\text{min}})/2}\cos\left(\int_{x_{\text{min}}}^x\frac{m e^x}{\mathcal{H}_{\Lambda\text{CDM}}}\dd{x}\right),
\end{equation}
valid for $x>x_{\text{min}}$.
\begin{figure}
	\includegraphics[width=\linewidth]{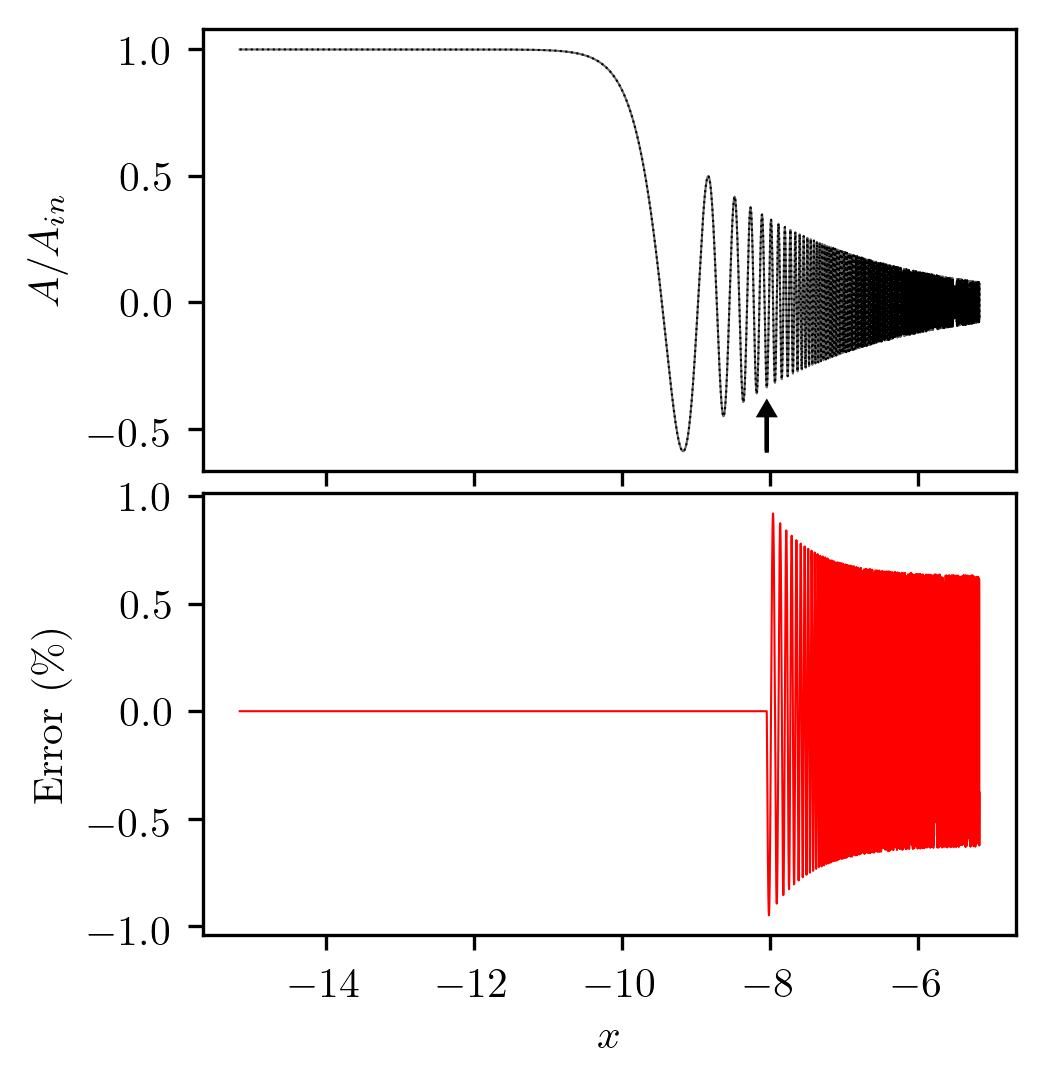}
	\caption{Comparison between the exact solution and our two-part model, employing the WKB approximation, for $m=10^{-26}$ eV. The upper panel shows the exact solution (solid line) and the approximate one (dotted line), displaying a very small difference, as well as an arrow that indicates the minimum from which the WKB solution is used. The lower panel shows the relative error, compared to the oscillation amplitude, which remains below $1\%$. The oscillatory behaviour indicates that the error is mostly due to a small phase difference between both solutions.}
	\label{fig:wkberror}
\end{figure}

With this solution, we can calculate the abundance today and get the primordial value $A_{\text{in}}$ in terms of $\Omega_A$
\begin{equation}
	A_{\text{in}}=\left(\frac{3\Omega_A H_0^2}{4\pi G m^2 e^{x_{\text{min}}}A_{\text{min}}}\right)^{1/2},
\end{equation}
thus fixing the boundary problem issue. Fig. \ref{fig:wkberror} shows a comparison between an exact solution and the solution we will be using, thus leading to small errors, and even less in a differential equation integration due to the oscillatory behaviour. Being able to use the later allows us to compute results much faster, as the large amount of oscillations before reaching the present time would have taken much computation time.

This two-part solution for the vector field can be plugged into the complete Hubble parameter \eqref{eq:H2x}, where, in the combination $A'=\mathcal{H}\partial_x A$, we must use $\mathcal{H}_{\Lambda\text{CDM}}$ in order to have a consistent solution, as this is the Hubble parameter that we have used to solve $A$. Finally we solve the GW evolution \eqref{eq:hl2x}. The starting integration point $x_{\text{in}}$ must be early enough so that the field is still constant and the GW mode is super-Hubble with $\mathcal{H}(x_{\text{in}})\gg \{e^{x_{\text{in}}}m, k\}$.

\begin{figure}
	\includegraphics[width=\linewidth]{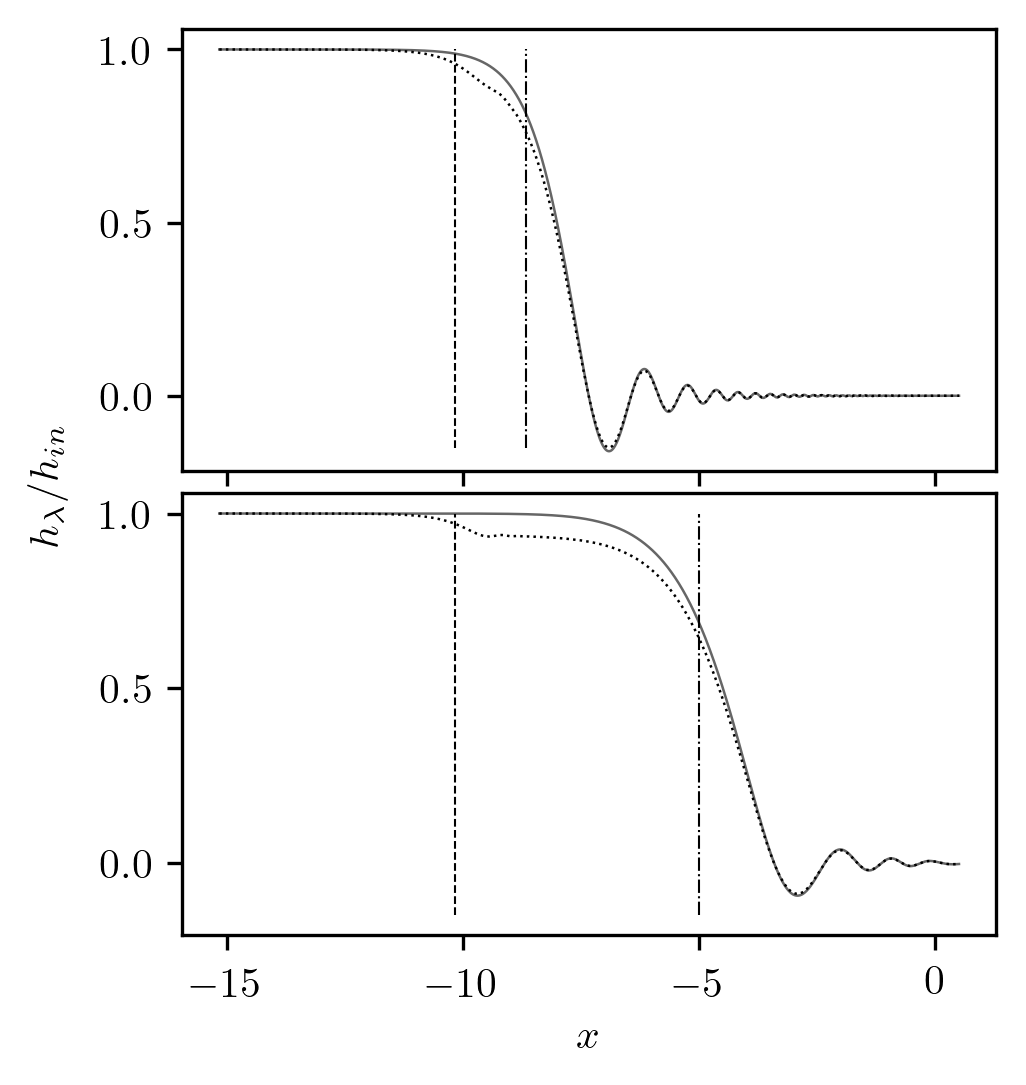}
	\caption{GW evolution for a $\Lambda$CDM background and for our model, in solid and doted lines respectively, and wavenumbers of $k=10^{-31}$ eV (above) and $k=10^{-32}$ eV (below). The rest of the parameters are fixed to $m=10^{-26}$ eV, $\Omega_A=0.25$, $\theta=\pi/3$. Vertical lines correspond to the ULVF oscillation beginning $ma=\mathcal{H}$ (dashed) and the GW entering the Hubble horizon $k=\mathcal{H}$ (dash-dotted).}
	\label{fig:hcompar}
\end{figure}

The GW propagation equation \eqref{eq:hl2x} must be solved twice, once for a $\Lambda$CDM background ($\Omega_A=0$) and then again for our model with a certain set of parameters. Fig. \ref{fig:hcompar} shows a couple of examples of GW propagation with and without the vector field. The main effect, which is fairly visible in those figures, is that the GW amplitude experiences a smooth diminution the moment the ULVF starts behaving as matter as long as this event occurs whilst the GW mode is super-Hubble. If it occurs after the wave entering the Hubble horizon, little to no effect is noticed.

\begin{figure}
	\includegraphics[width=\linewidth]{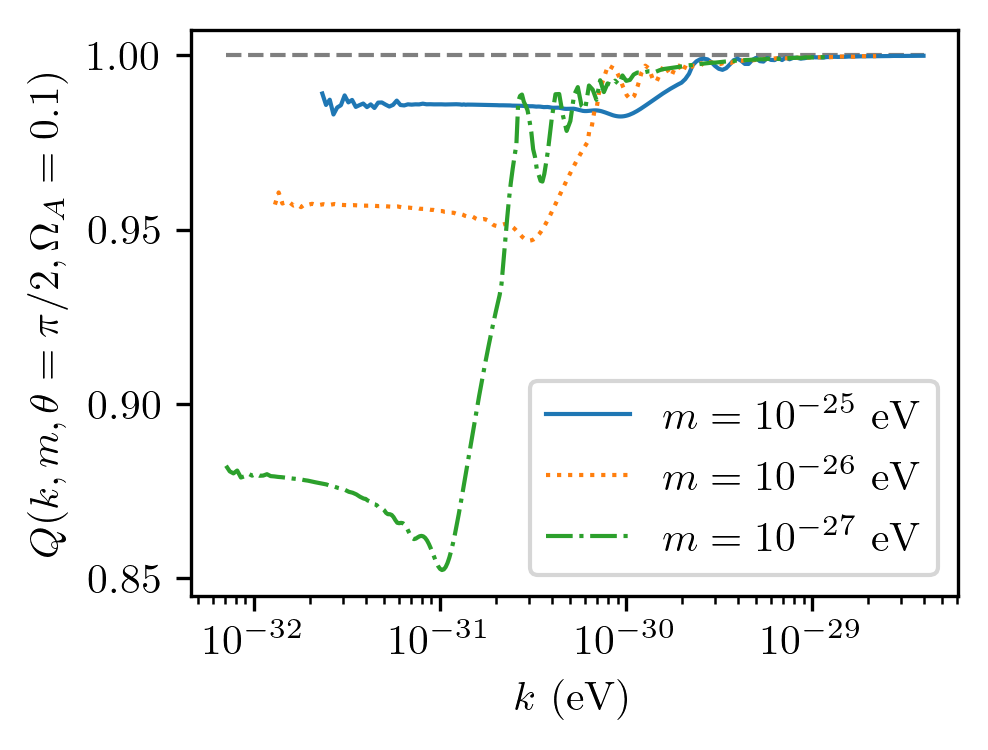}
	\caption{Quotient function for different ULVF masses, with fixed $\theta=\pi/2$ and $\Omega_A=0.1$. The deviation from $\Lambda$CDM $Q=1$ is larger the smaller the mass is. There is a global minimum at $k_{\text{min}}$, different for each mass, where the deviation is maximum. For $k\ll k_{\text{min}}$, the quotient tends to a constant value smaller than 1 (with little deviations in the figure due to numerical inaccuracy), whereas for $k\gg k_{\text{min}}$ it tends to unity.}
	\label{fig:Qm}
\end{figure}

\begin{figure}
	\includegraphics[width=\linewidth]{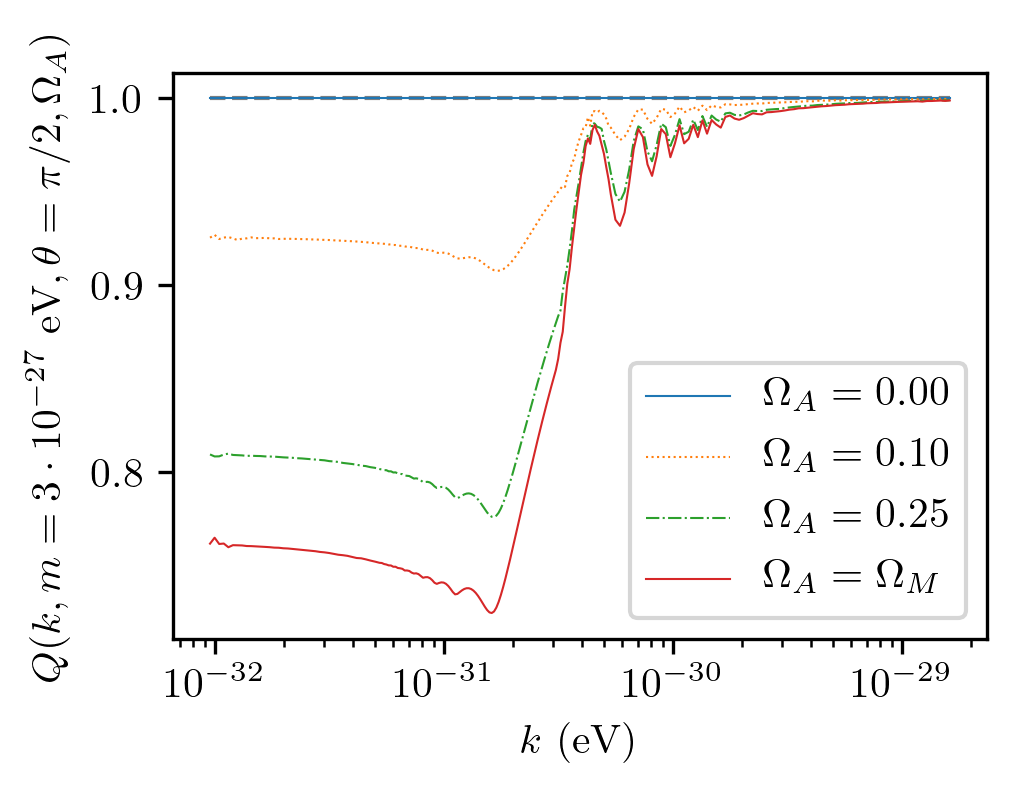}
	\caption{Quotient function for different ULVF abundances, with fixed $m=3\cdot 10^{-27}$ eV and $\theta=\pi/2$. The deviation from $\Lambda$CDM is larger the greater the abundance, which has a natural bound at $\Omega_M$. The wavenumber of maximum deviation (global minimum) does not depend on $\Omega_A$, as opposed to mass (see Fig. \ref{fig:Qm}). The choice $\Omega_A=0$ is exactly $\Lambda$CDM, as expected.}
	\label{fig:Qwa}
\end{figure}

\begin{figure}
	\includegraphics[width=\linewidth]{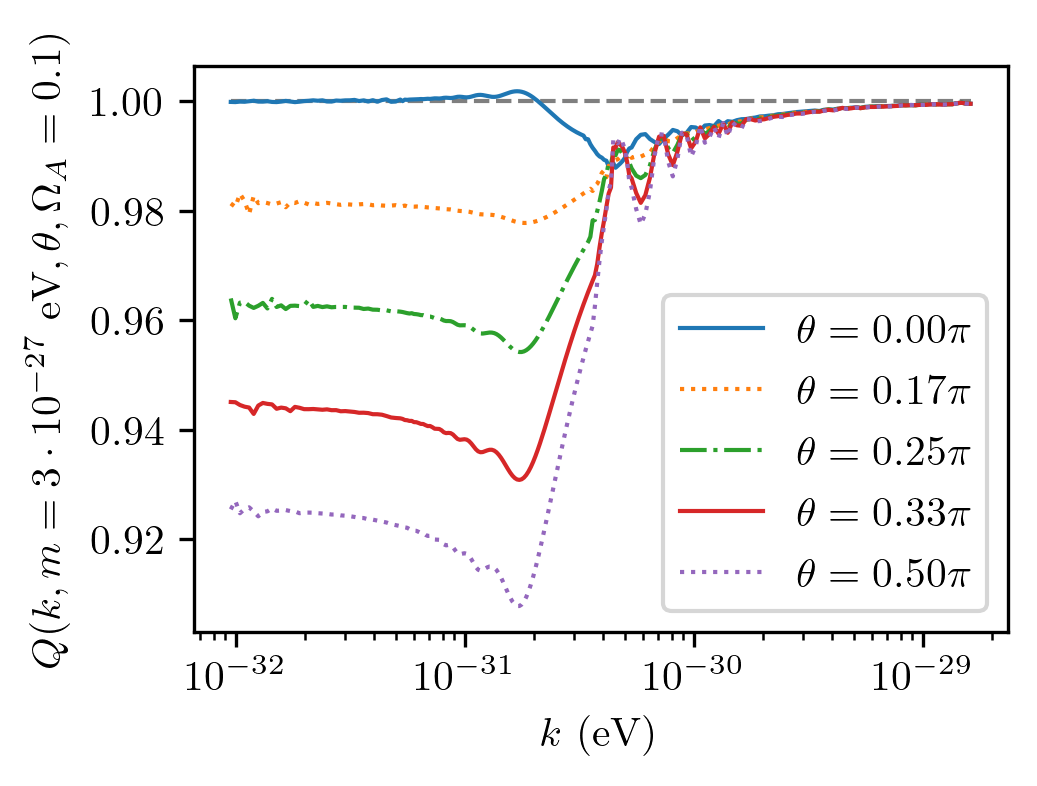}
	\caption{Quotient function for different vector field-wavevector angles $\theta$, with fixed $m=3\cdot 10^{-27}$ eV and $\Omega_A=0.1$. The deviation from $\Lambda$CDM is larger the closer to $\pi/2$ $\theta$ is. Angles in the range $\pi/2<\theta\leq\pi$ yield results that can be obtained from this picture, as the dependence in the propagation equation \eqref{eq:h2eq} is $\sin^2\theta$. Note that the wavenumber of maximum deviation does not depend on $\theta$ (obviating the extreme choice $\theta=0$), and that the case $\theta=0$ is not exactly $\Lambda$CDM despite having no extra term in the propagation equation, as the ULVF still affects the background evolution.}
	\label{fig:Qth}
\end{figure}

\begin{figure}
	\includegraphics[width=\linewidth]{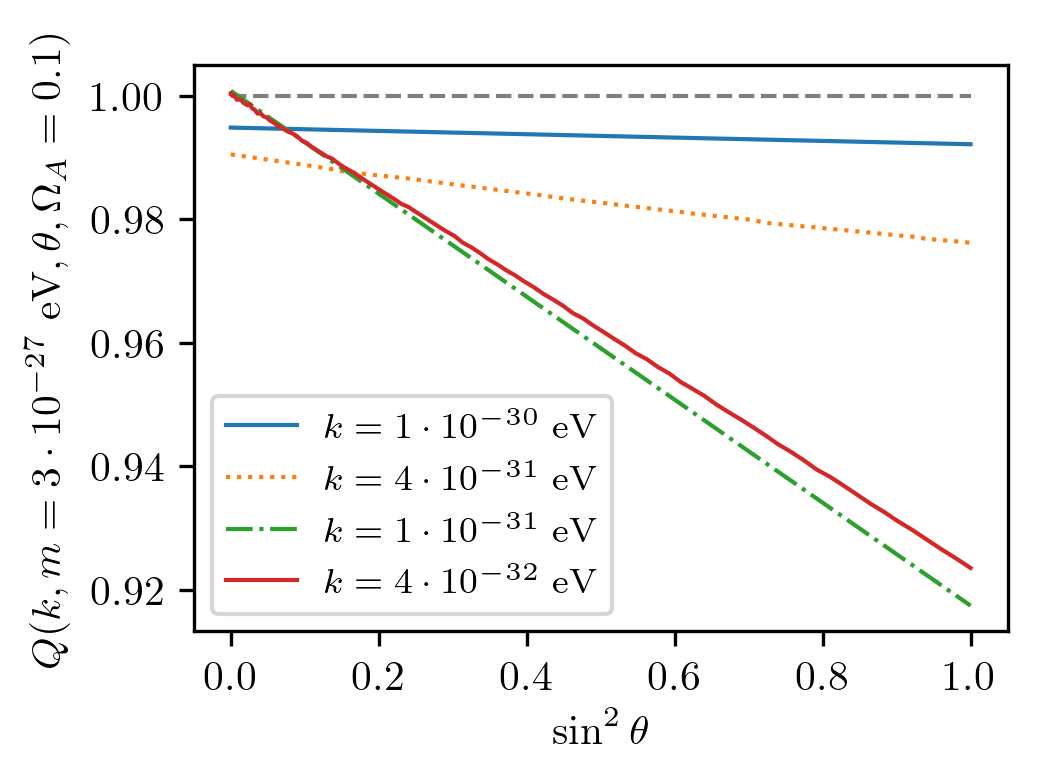}
	\caption{Quotient function as a function of $\sin^2\theta$, with fixed $m=3\cdot 10^{-27}$ eV and $\Omega_A=0.1$, for various wavenumber values. The dashed grey line corresponds to $\Lambda$CDM. The almost linear dependence suggests  an approximately  quadrupolar angular  modulation of the GW power spectrum.}
	\label{fig:Qkth2}
\end{figure}

With both solutions at hand, the transfer function \eqref{eq:Tfun} can be obtained from the $\Lambda$CDM one by directly dividing the value of the GW today by its primordial value. The quotient function \eqref{eq:Qfun} is computed through the comparison of both solutions for sufficiently long times, as discussed in Section \ref{sec:perturbations}. An example for different masses can be seen in Fig \ref{fig:Qm}, showing that there is a specific region where the effects are more noticeable, which corresponds to wavenumbers around $k=H_0/\sqrt{a_*}$. As $a_*$, the scale factor at which the ULVF starts behaving as matter, depends solely on the ULVF mass, the $k$ value for which the quotient function reaches its absolute minimum depends only on the mass as well, meaning that the variation of the angle $\theta$ or $\Omega_A$ produces a shift in the vertical axis, but not in the horizontal one. In particular, the deviation from $\Lambda$CDM is bigger the larger $\Omega_A$ is and the closer to $\pi/2$ the angle $\theta$ is, as can be seen in Figs. \ref{fig:Qwa} and \ref{fig:Qth} respectively. One last point to note is that setting $\Omega_A=0$ in our model corresponds exactly with $\Lambda$CDM, thus resulting in unity quotient function, but setting $\theta=0$ does not. This is because, despite the fact that $\theta=0$ eliminates the additional term in \eqref{eq:hl2x}, there is still a slight effect in the background accounted in $\mathcal{H}$, so we get $Q\neq 1$ even though there is no direct coupling between GWs and the ULVF in the propagation equation.

The quotient function exhibits non-trivial $k$ and $\theta$ dependencies as shown in Figs. \ref{fig:Qm} and \ref{fig:Qkth2}, which could make it possible to detect the presence of ULVF dark matter
in certain mass ranges.  Forthcoming experiments, both ground-based such as BICEP Array \cite{Hui:2018cvg} or Simons Array \cite{Suzuki:2015zzg}, and satellite-based like LiteBIRD \cite{Hazumi:2019lys}, are expected to detect primordial gravitational waves with a sensitivity of $\sigma(r)<0.006$ in the tensor to scalar ratio (even smaller in the case of LiteBIRD $\sigma(r)<0.001$), enough to detect typical values predicted by inflation around $r=0.01$, through CMB B-mode observations for $\ell<200$, as larger multipoles are dominated by gravitational lensing. Therefore, some effects in the B-mode power spectrum resulting from the ULVF presence are expected, on the one hand, as an angular modulation of the tensor power spectrum in that multipole range. As shown in Fig. \ref{fig:Qkth2}, this modulation would be approximately quadrupolar. On the other hand, we also expect a modification in the power-spectrum $k$ dependence. Both effects would more important for masses near the lower limit of $m\sim10^{-27}$ eV, as can be seen in Fig. \ref{fig:Qm}. Notice also that astrophysically-generated GWs typically have larger frequencies, where the quotient function tends to unity and therefore no effect in propagation is expected.  

\begin{figure}
	\includegraphics[width=\linewidth]{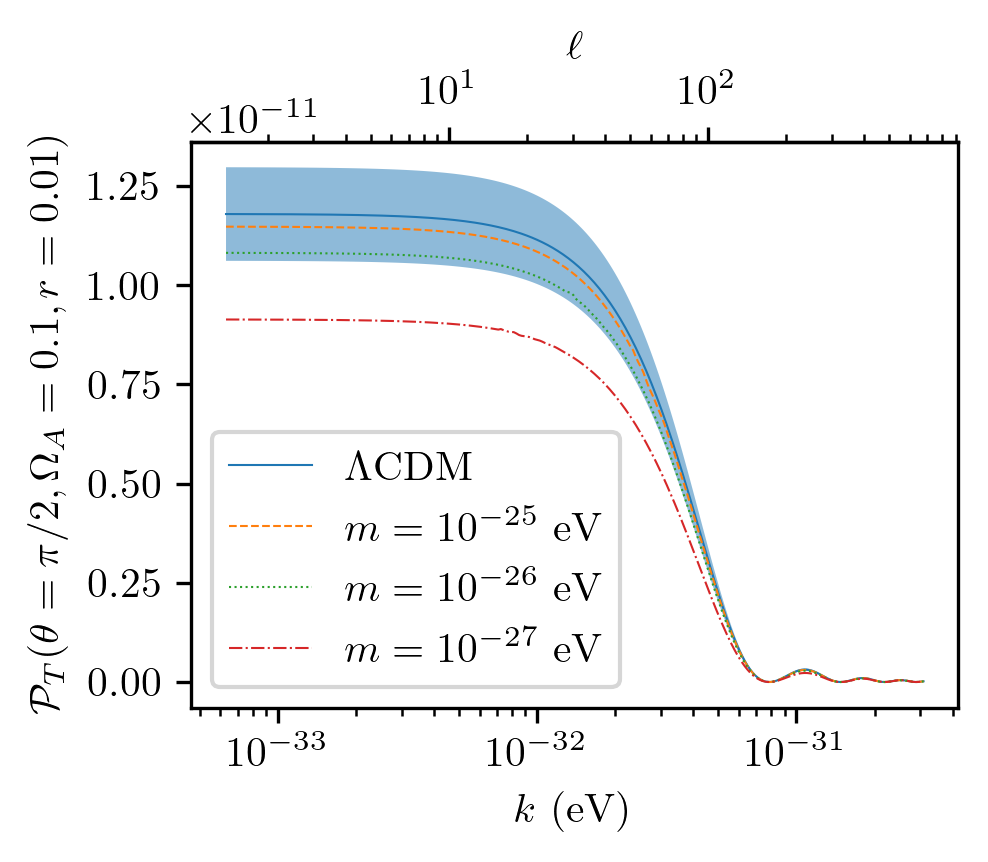}
	\caption{Tensor power spectrum at decoupling time for different ULVF masses, $\theta=\pi/2$, ULVF abundance of $\Omega_A=0.1$, primordial  tensor-to-scalar ratio $r=0.01$ and no spectral tilt $n_T=0$. The rest of parameters in Eq. \eqref{eq:omegagw2}, corresponding to the scalar sector of the primordial power spectrum correspond to  the Planck cosmology \cite{Aghanim:2018eyx}. The upper abscissa axis represents the approximate multipole moment $\ell$ corresponding to the wavenumber, taking into account its angular size at the last scattering surface. The blue band represents the precision of $\sigma(r)=0.001$ of LiteBIRD. Smaller masses (as well as bigger ULVF abundances) are more likely to be detected as deviations from $\Lambda$CDM.}
	\label{fig:Omegam}
\end{figure}

We plot in Fig. \ref{fig:Omegam} the tensor power spectrum at decoupling time  for various sets of parameters, in which we can see that the ULVF presence suppresses the power with respect to  $\Lambda$CDM  for small masses. The effect is more evident at low $k$ (multipoles $\ell<100$), where the quotient function is not unity for these small masses. Such low frequency GW  are far from being directly observable with interferometers or pulsar timing arrays taking into account current or planned detector sensitivities and wavelength ranges \cite{Schmitz:2020syl}, so the observation of these effects is constrained to the measurement of B modes. 

\section{Conclusions}\label{sec:conclusions}

Ultralight vector fields are suitable dark matter candidates. While contributing to the observed and yet unexplained dark matter abundance today, their coherent oscillatory behaviour
produces interesting new phenomenology compared to standard cold dark matter. 

We have found that primordial GWs, those coming from the inflationary epoch, see their amplitudes slightly diminished for small ULVF masses $m$ and large abundances $\Omega_A$, as long as the GW mode enters the Hubble horizon after the ULVF starts to behave as matter, which occurs for long-wavelength modes. This suppression is anisotropic and is maximized for GWs propagating orthogonally to the direction of the background vector field. In addition, as can be seen in Fig. \ref{fig:Qm}, there is a wavenumber region for each mass in which there is a non-zero change with respect to $\Lambda$CDM. Therefore, the presence of the ULVF is expected to modify the CMB B-mode signal, either as an angular modulation or a change in the shape of the power spectrum. This is particularly interesting  in the low-multipole ($\ell\leq 200$) region, where gravitational lensing is not dominant. Taking into account the sensitivity and multipole region of forthcoming observations, we infer from the tensor power spectrum at decoupling time (Fig. \ref{fig:Omegam}) that the effects of ULVF dark matter could be significant enough to be detectable for masses larger than the lower bound of $m\sim 10^{-27}$ eV and smaller than an upper bound of $m\lsim 10^{-26}$ eV. This range of mass has an associated de Broglie wavelength of astrophysical size, and as a result it is not suitable for describing the totality of the dark matter abundance but a part of it, as commented previously in Section \ref{sec:Intro}. We have also seen that the effect on the propagation of astrophysically-generated GW is completely negligible.

A more detailed study of these effects would require the computation of the complete B-mode power spectrum, which could be obtained by the modification of any of the already existing Boltzmann codes, such as CLASS \cite{Blas:2011rf}, in order to include the ULVF in the background and metric perturbation equations. With the B-mode power spectrum at hand, one could conclude if the deviations from $\Lambda$CDM that we have predicted in this work are large enough to be distinguishable in CMB measurements by direct comparison. At the moment, this task is beyond the scope of our work. It is also interesting to explore the possibilities to generate the ULVF dark matter abundance by different suitable mechanisms such as misalignment \cite{Nelson:2011sf} whilst avoiding the suppression during inflation. Finally, the 
effects on GW propagation in vector theories with different potentials could also provide a rich phenomenology.

\vspace{1cm}

\acknowledgements{This work has been supported by the MINECO (Spain) projects FIS2016-78859-P and PID2019-107394GB-I00 (AEI/FEDER, UE). A.D.M. acknowledges financial support by the MICIU (Spain) through a Formación de Profesorado Universitario (FPU) fellowship FPU18/04599.}

\bibliography{paper-ulvfmass}

	\end{document}